\documentclass{aastex}
\usepackage{epsfig}
\begin{document}
\bibliographystyle{plainnat}
\voffset-1cm
\newcommand{\gsim}{\hbox{\rlap{$^>$}$_\sim$}}
\newcommand{\lsim}{\hbox{\rlap{$^<$}$_\sim$}}

 \title{Kinematic Origin Of Correlations\\
       Between Gamma Ray Burst Observables} 

\author{Shlomo Dado\altaffilmark{1} and Arnon Dar\altaffilmark{2}}

\altaffiltext{1}{dado@phep3.technion.ac.il\\
Physics Department, Technion, Haifa 32000,
Israel}
\altaffiltext{2}{arnon@physics.technion.ac.il\\
Physics Department, Technion, Haifa 32000, Israel}

\begin{abstract}

Recently, several new correlations between gamma ray burst (GRB) 
observables have been discovered. Like previously well established 
correlations, they challenge GRB models. Here we show that in the 
cannonball (CB) model of GRBs, the newly discovered correlations have the 
same simple kinematic origin as those discovered earlier. They all result 
from the strong dependence of the observed radiations on the Lorentz 
and Doppler factors of the jet of highly relativistic plasmoids 
(CBs) that produces the observed radiations by interaction with the 
medium through which it propagates.

\end{abstract}


\section{Introduction} 

Despite the enormous complexity and diversity of gamma ray bursts (GRBs) 
and their afterglows, various well established correlations between GRB 
observables were found during the past years. Such correlations challenge 
the theoretical models of GRBs. Most of these correlations have been 
neither predicted nor explained by the standard fireball model of GRBs, 
which has been extensively employed to explain the GRB phenomenon 
\citep[for reviews see, e.g.,][]{Meszaros2002, Zhang2004, Piran2004, 
Zhang2007}. In the cannonball (CB) model of GRBs, these correlations were 
shown to be a simple consequence of the strong dependence of GRB 
observables on the Lorentz factor $\gamma$ and Doppler factor $\delta$ of 
the highly relativistic jet of plasmoids (CBs) whose interaction with the 
medium along its path produces the observed radiations (Dar \& De 
R\'ujula~2000,2004; Dado et al.~2007,2009).

Recently, many new correlations were discovered between pairs of 
observables characterizing the prompt gamma ray and early optical 
emissions in a sample of GRBs rich in X-ray flashes \citep{Liang2010}. 
Here we show that in the CB model all these new correlations also follow 
from the strong dependence of the GRB observables on the Lorentz factor 
$\gamma$ and the Doppler factor $\delta$  of the 
highly relativistic jet of CBs.

In particular, from their selected GRB sample, Liang et al.~(2010) 
inferred tight 
pair correlations between the initial Lorentz factor $\Gamma_0$ of the 
relativistic ejecta in GRBs and other GRB observables, such as the 
isotropic equivalent gamma ray energy $E_{iso,\gamma}$ of the prompt 
emission, the peak luminosity $L_{p,O}$ of the early optical emission, and 
the peak time $t'_p$ of the optical emission in the GRB rest frame. The 
standard fireball model with internal and external shocks 
does not explain the physical origin of these newly 
discovered correlations, nor the origin of other well established 
correlations between GRB observables.

Moreover, in the internal shock fireball model, the 
tight $\Gamma_0-E_{iso, \gamma}$ correlation yields a peak energy 
$E'_{p,\gamma}$ that is practically independent of $E_{iso,\gamma}$
\citep{Liang2010}, in 
contradiction with the well established Amati relation\footnote{
See, however, Thompson et al.~2007, and Zhang \& Yan~2011.}
\citep{Amati2002,Amati2006}. However, the initial value of the Lorentz 
factor of the relativistic ejecta that produced the GRBs was not truly 
measured. Instead, it was inferred \citep{Liang2010} from an assumed 
relation, $\Gamma_0\propto [E_{iso,\gamma}/(t'_p)^3]^{1/8}$. This relation 
follows from the standard fireball model expression for $\Gamma_0$ 
\citep{Sari1999}, neglecting its weak dependence on the circumburst 
density and on the conversion efficiency of the relativistic kinetic 
energy of the ejecta to radiation. The validity of this fireball model 
relation has never been proven. Thus, hereafter, we shall distinguish 
between 
the true initial value of the bulk motion Lorentz factor $\gamma$ of the 
relativistic ejecta and $\Gamma_0$, its alleged value inferred from 
fireball modeling of GRB data.  We shall show that, in the CB model, the 
strong dependence of $E_{iso,\gamma}$ and $t'_p$ and consequently of 
$\Gamma_0$ on the true initial values of the Lorentz factor $\gamma$ and 
the Doppler factor $\delta$ also yields the tight pair correlations 
between 
the fireball model parameter $\Gamma_0$ and the GRB observables which were 
discovered by Liang et al.~(2010) and by Lu et al.~(2011), and 
could not be explained by the standard
fireball model\footnote{see, however, Lu et al 2011.}.
 
\section{Kinematic origin of correlations in the CB model} 

\subsection{Origin of the observed radiations in the CB model}

In the cannonball (CB) model of GRBs \citep{DDD2002, DD2004, DDD2009}
GRBs and their afterglows are produced by the interaction of bipolar 
jets of highly relativistic plasmoids (CBs) of ordinary matter with the 
radiation and matter along their trajectory \citep{SD1995, Dar1998}. Such 
jetted CBs are presumably ejected in accretion episodes on the newly 
formed compact stellar object in core-collapse supernova (SN) explosions 
\citep{Dar1992,DP1999,DD2000}, in merger of compact objects in close 
binary systems \citep{Goodman1987,SD1995} and in phase transitions in 
compact stars \citep{Dar1998, DD2000, DDD2009}. For instance, in  
GRBs associated with SNe,
it is hypothesized that an accretion disk or a torus is produced around 
the newly formed compact object, either by stellar material originally 
close to the surface of the imploding core and left behind by the 
explosion-generating outgoing shock, or by more distant stellar matter 
falling back after its passage \citep{DD2004}. As observed in 
microquasars, each time part of the accretion disk falls abruptly onto the 
compact object, two CBs made of ordinary-matter plasma
are emitted with large bulk-motion Lorentz factors $\gamma\!\gg\!1$ 
in opposite directions along the rotation axis from where matter has 
already fallen back onto the compact object due to lack of rotational 
support. The prompt $\gamma$-ray pulses and early-time and X-ray flares 
are dominated by inverse Compton scattering (ICS) of glory photons - a 
light halo surrounding the progenitor star that was formed by scattered 
stellar light from the pre-supernova wind blown from the progenitor star. 
The ICS is overtaken by synchrotron radiation (SR) when the CBs enter the 
pre-supernova wind/ejecta of the progenitor star (see, e.g., Dado et 
al.~2009). The SR dominates the early time optical/NIR emission and the 
broad-band afterglows produced by the CBs when they continue to propagate 
in the interstellar medium. ICS of the SR produces the emission of very 
high energy photons during the early time optical/NIR emission and the 
broad band afterglow \citep{DD2009a}.

\subsection{Kinematic correlations} 
GRBs are not standard 
candles because of the 
diversity of their central engines and environments. But, because of the 
large bulk motion Lorentz factor $\gamma$ of the jet of CBs, their 
emitted  radiation 
at redshift $z$, which is observed at a small angle $\theta$ relative to 
the direction of the jet, is boosted 
by a large Doppler factor $\delta\!=\!1/\gamma\,(1\!-\!\beta\,cos\theta)$
and collimated through relativistic beaming by a factor $\delta^2$.
Moreover, the time difference $dt$ in the observer frame between the 
arrival of photons emitted by the point-like CBs at two different points 
along their path, which are separated by a distance $dr'$ and time 
$dt'\!=\!dr'/c$ in the progenitor's rest frame (hereafter a prime 
indicates an observable in the progenitor's rest frame), is shortened 
(aberrated) according to $dt\!=\!dt'\,(1\!+\!z)/\gamma\, \delta$. The 
large Doppler boosting, relativistic collimation and time aberration 
produce correlations between GRB observables, despite their dependence on 
the CBs' intrinsic (rest frame) properties and on the environment along 
their trajectories (which produce a significant spread around these 
simple kinematic correlations).

The redshift $z$ of the GRB location is measurable, and the dependence of 
the GRB
observables on redshift can be taken into account explicitly,
unlike their dependence on the values of the Lorentz factor and the
viewing angle of the jet, which can only be inferred with model-dependent
assumptions. However, the strong dependence on $\gamma$ and $\delta$
can be used to
correlate triplets of independent observables without knowing the
values of $\gamma$ and $\delta$. Moreover, several observables
depend on the same combination of $\gamma$ and $\delta$ that result in
pair correlations. Finally, due to selection effects in the observations,
various observables depend strongly only on $\gamma$ or $\delta$,
which also yields pair-correlations. 
In particular,
the dependence on viewing angle of the jet can be eliminated in two 
general cases:
For $\gamma^2\!\gg\! 1$ and small viewing angles 
$\theta^2\!\ll\! 1$, the Doppler factor satisfies
$\delta\! \approx\! 2\, \gamma/(1\!+\!\gamma^2\, \theta^2)$ 
to an excellent approximation. 
For  $\theta^2\,\gamma^2\lsim 1$ the Doppler factor decreases 
rather slowly with increasing $\theta$. But, 
for $\theta^2\,\gamma^2\!\gg\! 1$ the Doppler factor decreases with 
increasing viewing angle like $\theta^{-2}$, and the observed fluence of 
gamma rays, which in the CB model is amplified by a factor 
$ \delta^2 $ due to relativistic beaming, decreases
like $[1+\gamma^2\theta^2]^{-2}$. The geometrical probability to view a 
bipolar GRB from 
a small angle $\theta$ increases like $(1\!-\!cos\theta)\!\approx\! 
\theta^2/2$ and the product $\theta^2\,[1+\gamma^2\theta^2]^{-2}$
has a maximum when $\gamma^2\theta^2\!=\!1$. 
Consequently, $\delta\!=\!\gamma$ for the most probable viewing angle 
$\theta\!=\!1/\gamma$  of GRBs \citep{SD1995}.  

In `soft' GRBs with $\gamma^2\,\theta^2\!\gg\!1$, such as X-ray flashes 
(XRFs) with $\gamma^2\,\theta^2 \gsim 10$, the dependence on the exact 
value of the Lorentz factor can be ignored  compared  to the strong 
dependence on the Doppler factor. This yields a slightly 
different correlation. Thus, we shall derive the CB model correlations for 
the above two different situations, for $\delta\!\simeq\!\gamma$
and for $\gamma^2\theta^2\!\gg\!1$, i.e. $\delta\!\ll\!\gamma$
where the dependence on $\gamma$ can be neglected compared to that on 
$\delta$. 
For a mixed 
population of ordinary GRBs, soft GRBs and XRFs one may expect an 
approximate 
correlation obtained by using the average index of the two power-law 
correlations. (Alternatively, one can derive triple correlations, i.e., 
correlations that involve three independent observables that each of
which depends both on $\gamma$ and $\theta$). Below we derive the
the CB model pair correlations that corresponding to those
discovered empirically  by  Liang et al.~(2010) and Lu et al.~(20111) 
and we compare them  in the text and in  Table 1.

\section{Pair Correlations between GRB observables}

\subsection{Correlations between pulse-shape parameters}

In the CB model, $\gamma$  and $\delta$ stay put at their 
initial values until the CB sweeps in a relativistic mass/energy 
comparable to its initial rest mass, e.g., during the prompt emission 
and the early afterglow until the 'jet break'  
(Dado et al 2009 and references therein), or
during flares due to crossing of density bumps with a wind-like 
density profile.    
Because of time-aberration, all time measures of the prompt
emission pulses and early time flares
such as their rise-time $t_r$ from 
half maximum to peak value, the peak-time $t_p$
{\it after the beginning of the flare/pulse}, the
decay-time $t_d$ from peak value to half maximum,
and its full width at half
maximum (FWHM)  $t_w$, are proportional to 
$(1\!+\!z)/\gamma\,\delta$.
Hence,
\begin{equation}
log t_i=a_{ij}+log t_j\,, {\rm where}\,\, i,j=r,d,p,w.
\label{tcorel}
\end{equation}
This explains the origin of the 'universal' power-law index $\!\sim\!1$
of the `power-law correlations' (proportionality) between the temporal
parameters of the prompt $\gamma$-ray pulses.

Moreover, the prompt ICS emission pulses and early time flares
have an approximate light curve (see, e.g., Dado et al.~2009),
\begin{equation}
E\, {d^2N_\gamma\over dt\,dE}(E,t)
\propto {t^2 \over(t^2+\Delta^2)^2}\,E{dN_\gamma\over dE}
\label{pulseshape}
\end{equation}
where $\Delta$ is the characteristic time before the fast expansion of a 
CB, roughly like $R^2 \propto t^2/(t^2+\Delta^2)$,
stops by cooling or by the collision with the wind ejecta,
which, like the glory, has roughly $1/(r^2+r_g^2)\propto 1/(t^2+\Delta^2)$ 
decline. As long as the
spectral evolution is slow during a pulse/flare, the pulse shape can be 
well approximated by $E{dN_\gamma/dt}\propto t^2/(t^2+\Delta^2)^2$,
which peaks at $t_p\!=\!\Delta$, has a full width at half maximum
$t_w\!=\!2\,\Delta$, a rise time from half maximum to peak value,
$t_r\!\approx\!0.59\,\Delta$ and a decay time from peak value to half 
maximum $t_d\!\approx\!2.84\Delta$. Consequently, 
the  above 'universal' pulse/flare shape 
yields the simple redshift independent relations  \citep{DD2004, 
DDD2009}:\\
$t_d\!\approx\!2.39\,t_r$, i.e., $log\,t_d\!\approx\!0.38\!+\!log\,t_r$,\\
$t_d\!\approx\!1.41\,t_p$, i.e., $log\,t_d\!\approx\!0.15\!+\!log\,t_p$,\\ 
$t_r\!\approx\!0.59\,t_p$, i.e., $log\,t_r\!\approx\!-0.23\!+\!log\,t_p$,\\  
$t_w\!\approx\!2.00\,t_p$, i.e., $log\,t_w\!\approx\!0.30\!+\!log\,t_p$,\\  
$t_w\!\approx\!3.39\,t_r$, i.e., $log\,t_w\!\approx\!0.53\!+\!logt\,_r$,\\
$t_w\!\approx\!1.42\,t_d$, i.e., $log\,t_w\!\approx\!0.15\!+\!log\,t_d$,\\  
where time is measured in seconds. These  relations 
are well satisfied within observational errors by well resolved GRB 
peaks/flares as was found, e.g., in \citep{Kocevski2003} through 
empirical parametrization of  these peaks/flares.
In the CB model, these correlations are valid approximately also for the 
prompt/early time optical flares 
because both the prompt/early time ICS pulses and the 
prommpt/early time ICS and SR flares 
rise 
and a decay like $(\Delta^2+t^2)^{-1}$.
Indeed, the above predicted correlations are in good agreement with those 
found by Liang et al (2010) for the early time optical peak:\\ 
$log\,t_d = (0.48\pm 0.13) + (1.06\pm 0.06)\, log\, t_r$,\\ 
$log\,t_d = (-0.09\pm 0.29) + (1.17\pm 0.11)\, log\, t_p$,\\ 
$log\,t_r = (-0.54\pm 0.22) + (1.11\pm 0.08)\, log\, t_p$,\\
$log\,t_w = (0.05\pm 0.27) + (1.16\pm 0.10)\, log\, t_p$,\\ 
$log\,t_w = (0.61\pm 0.11) + (1.05\pm 0.05)\, log\, t_r$,\\ 
$log\,t_w = (0.15\pm 0.02) + (0.98\pm 0.01)\, log\, t_d$.\\ 
However Liang et al.~(2010) have  selected 
for their analysis GRBs with a single optical peak that could be 
described 
by the empirical formula of Kocevski et al.~(2003) for GRB pulses.
In most GRBs, however, the prompt optical emission is probably a sum of 
unresolved overlapping flares, as suggested by several bright GRBs such as 
050820A \citep{Vestrand2006}, 080319B \citep{Wozniak2009}, 061007 
\citep{Rykoff2009} and 071031 \citep{Kruhler2009}, where the large photon 
statistics allowed a good temporal resolution with rapid-response 
ground-based telescopes.  In nearby soft GRBs (GRBs with a relatively 
small $E'_p$) and XRFs, the individual flares are streched in time 
because of a much smaller Doppler factor, while the duration of the 
central engine activity remains the same, independent of viewing angle. 
Consequently, in far off-axis GRBs, such as XRFs, the individual flares 
strongly overlap and produce an effective single peak. Such peaks are 
hardly resolved into overlapping flares even with large rapid-response 
telescopes such as GROND \citep{DD2010}. Thus, the sample of GRBs in 
Liang et al.~(2010) that consists mainly of soft GRBs and XRFs have a 
single optical peak at a relatively large peak-time, which allowed their 
measurement with good temporal resolution with rapid response telescopes 
that are larger but slower than the  robotic telescopes. 

Note also that because the peak energy $E_{p,\gamma}$
of the prompt $\gamma$-ray pulses and of the early-time  X-ray flares 
is proportional to $\gamma\,\delta/(1\!+\!z)$ \citep{DD2004,DDD2009}, 
all the above temporal parameters 
of the prompt $\gamma$-ray, X-ray and optical peaks/flares 
are also inversely proportional to $E_{p,\gamma}$ of the $\gamma$-ray 
emission,
$t_r\!\propto\!1/E_{p,\gamma}$,
$t_p\!\propto\!1/E_{p,\gamma}$,
$t_d\!\propto\!1/E_{p,\gamma}$, 
$t_d\!\propto\!1/E_{p,\gamma}$.
Both, the temporal correlations and these correlations are independent of 
redshift.

\subsection{Pair correlations among optical emission observables}

In the CB model, the optical emission is dominated by synchrotron 
radiation (SR) and its lightcurve is 
given by \citep[see, e.g.,] [and references therein]{DDD2009},
\begin{equation}
F_\nu[t] \propto n^{(1+\beta)/2}\,R^2 \gamma^{3\, \beta-1}\delta 
^{3+\beta}\,, 
\label{SRlc}
\end{equation} 
where $n[t]$ is the density along the CB trajectory, $R$ is the CB radius 
and $\beta$ is the spectral index of the SR. Typically, the spectral index 
in the optical band  is 
$\beta_O\!\approx\! 0.5$   at early time and gradually approaches 
$\beta_X\!\approx\! 1$ at late time. 
The peak luminosity of the prompt
optical emission flare is obtained when the CB reaches the peak density 
of the progenitor's wind/ejecta \citep{DD2010} while $\gamma$ 
and $\delta$ 
stay put at their initial values.
Thus, for the most probable  viewing angle of GRBs 
$\theta\!\approx\!1/\gamma$, i.e., $\delta\!\approx\!\gamma$, 
the CB model predicts for the peak optical luminosity 
$L_{p,O}\!\propto\!\gamma^4$, and  $t'_w \!\propto\! 1/ 
\gamma\,\delta\!\approx\! 1/\gamma^2$ for the FWHM in the GRB rest frame.
These dependencies  yield the correlation
$L_{p,O}\!\propto\! [t'_w]^{-2}$, 
in agreement with that found in \citep{Liang2010}
and reported in their equation (11) and in Table 1 below. 
Similar correlations are expected  between $L_{p,O}$
and $t_r$, $t_d$ and $t_p$, respectively .

Note, however, that 
in the CB model, XRFs that are far off-axis GRBs
have  initially $\gamma^2\,\theta^2\! \gg 1$. 
The deceleration of the CBs in XRFs yields $\delta(t)$, which 
first rises slowly as $\gamma(t)$ decreases with increasing time until it
reaches a maximum  when $\gamma(t)\theta\!=\!1$, i.e., when
$\delta(t)\!=\!\gamma(t)$. Hence,  
for a constant density ISM, 
a typical early-time optical spectral index $\beta_O\simeq 0.5$,
and $\delta(t)\!\simeq\! \gamma(t)$, Eq.~(\ref{SRlc}) yields
$L_{p,O}\propto [\gamma(t)]^4$. Hence, for 
$t'_w \propto 1/\gamma(t)\,\delta(t)\simeq [\gamma(t)]^{-2}$,
the CB model predicts the correlation $L_{p,O}\propto [t'_w]^{-2}$.

\subsection{Correlations involving total energy, peak energy and peak 
optical luminosity} 

In the CB model the  peak energy $E_{p,\gamma}$ and the
isotropic equivalent gamma ray energy $E_{iso,\gamma}$
of a {\it   single gamma-ray pulse or early-time flare} satisfy 
\citep{DD2000}
\begin{equation}
(1+z)\, E_{p,\gamma}\propto \gamma\,\delta, ~~~~E_{iso,\gamma}\propto 
\delta^3\, .
\label{Eg}
\end{equation} 
where $E_{iso,\gamma}$ and $E_{p,\gamma}$ refer to that single 
pulse/flare. 
For  $\delta\!\approx\!\gamma$, Eq.~(\ref{Eg}) yields 
$(1\!+\!z)\, E_{p.\gamma}\!\propto\! \gamma^2$, 
$E_{iso,\gamma}\!\propto\!\gamma^3$ and
the correlations 
$(1\!+\!z)\, E_{p,\gamma}\propto E_{iso,\gamma}^{2/3}$.
In soft GRBs and XRFs, 
i.e., GRBs with a large viewing angle, the CB model predicts 
 $(1\!+\!z)\, E_{p,\gamma}\!\propto\!\delta 
\!\propto\! E_{iso,\gamma}^{1/3}$ \citep{DD2000}.
For a mixed population of soft and hard GRBs, it yields the mean 
correlation 
\begin{equation} 
(1+z)\, E_{p,\gamma}\propto E_{iso,\gamma}^{1/2\pm 1/6},
\label{Eisomean}
\end{equation}
Since $E'_{p,\gamma}$ at peak luminosity  is observed to be proportional 
to $E'_{p,\gamma}$ of the time integrated spectrum over the entire 
GRB \citep{Goldstein2012}, the above $E'_{p,\gamma}-E_{iso,\gamma}$ 
correlation, is valid also for the entire 
GRB \citep{DD2000}, as  was discovered empirically  
\citep{Amati2002,Amati2006}.

In the CB model, where
the early optical emission is dominated by 
synchrotron radiation with a canonical spectral index $\beta_O\approx 
0.5$, Eq.~(\ref{SRlc}) predicts that  
ordinary GRBs with  $\delta\approx \gamma$ have  a peak optical 
luminosity $L_{p,O}\propto \gamma^{4}$ and consequently 
$L_{p,O}\!\propto\!E_{iso,\gamma}^{4/3}$. 
In soft GRBs and XRFs where the 
dependence on $\gamma$ at early time can be neglected compared to 
the dependence  on 
$\delta$, the resulting correlation is $L_{p,O}\!\propto\! 
\delta^{7/2}\!\propto\!E_{iso,\gamma}^{7/6}$. Thus, the mean correlation 
in a mixed population of soft and hard GRBs that is predicted by the CB 
model is, 
\begin{equation} L_{p,O} \propto E_{iso,\gamma}^{5/4\pm 1/12}\, ,
\label{Lisomean} 
\end{equation}
which is in agreement with the power-law 
correlation with an index $1.17\pm 0.13$ found by Liang et al.~(2010) and 
reported in their equation (12) and in Table 1 below.

The equivalent isotropic optical energy of the prompt optical flare
that dominates $E_{iso,O}$ is roughly given by $L_{p,O}\, t'_w 
\!\propto\! \gamma^{-1/2}\, \delta^{5/2}$.
Thus, in ordinary GRBs where $\delta\approx \gamma$, 
the CB model predicts $E_{iso,O}\!\propto\!\gamma^2 $, 
i.e., $E_{iso,O}\!\propto\! E_{iso,\gamma}^{2/3}$.
In  soft GRBs,  $E_{iso,O}\!\propto\!\delta^{5/2}\propto 
[E_{iso,\gamma}]^{5/6}$,  and
 the mean effective correlation that follows is,
\begin{equation}
E_{iso,O}\propto E_{iso,\gamma}^{3/4\pm 1/12}\,,
\label{lpoeiso}
\end{equation}
which is in good agreement with the correlation found by Liang et 
al.~(2010) and reported in their equation (14), and in Table 1 below .

\subsection{Triple correlations}

Many  correlations involving triplets of GRB observables can 
be derived using their strong dependence on $\gamma$
and $\delta$. For instance, 
the peak of the equivalent isotropic luminosity of a {\it single}  
$\gamma$-ray pulse or an early-time X-ray flare satisfies,
\begin{equation}
L_{p,\gamma}\approx E_{iso,\gamma}/t'_w \propto 
E_{iso,\gamma}\, E'_{p,\gamma}\,, 
\label{Lpg}
\end{equation}
where $L_{p,\gamma}, E_{iso,\gamma}$ and $E'_{p,\gamma}$ are of the 
same single pulse/flare. 

Since $E'_{p,\gamma}$ at peak luminosity is observed to be proportional to 
$E'_{p,\gamma}$ of the time-integrated spectrum over the entire GRB 
\citep{Goldstein2012}, the CB model yields the approximate binary 
correlation $E'_{p,\gamma}\propto L_{p,\gamma}^{1/3 \pm 1/9}$. This 
predicted correlation  is compared in Fig. 1 
with Fermi GBM observations of 26 GRBs with known redshift 
\citep{Gruber2011}.
Note, however, that the time-averaged luminosity of {\it multipeak} GRBs  
satisfies
\begin{equation}
<L_{iso, \gamma}>\approx 0.9\, \, E_{iso,\gamma}/ [T_{90}/(1+z)]\,,
\end{equation}
where $T_{90}/(1+z)$ is the  GRB duration 
during which $90\%$ of the observed prompt GRB energy is emitted.
In the CB model,  $T_{90}/(1+z)$ of multipeak long GRBs is intrinsic and 
does  not depend on $\gamma$ and/or $\delta$. Hence, the last equation    
yields the binary correlations
\begin{equation}
<L_{iso, \gamma}>\propto E_{iso,\gamma};\,\,\,     
 E'_{p,\gamma}\propto  <L_{iso, \gamma}>^{1/2 \pm 1/6}\, . 
\label{LE}
\end{equation}  

In the CB model, the break time of a canonical X-ray afterglow 
satisfies \citep{DDD2009}, $t_{b,X}\!\propto\! 
(1\!+\!z) / (\delta^2\, \gamma)$.
Then, using the relations in Eq.~(\ref{Eg}), the following 
correlation is obtained,  
\begin{equation}
t'_{b,X} \propto 1 / (E'_{p,\gamma}\, [E_{iso,\gamma}]^{1/3})\,, 
\label{tb}
\end{equation}
which yields the approximate binary correlations
\begin{equation}
t'_{b,X} \propto 1/[E'_{p,\gamma}]^{7/4 \pm 1/4}\propto  
1/[E_{iso,\gamma}]^{5/6 \pm 1/6}\,. 
\label{tbc}
\end{equation}

\section{Origin of the tight $\Gamma_0 - E_{iso,\gamma}$
and $\Gamma_0 - L_{iso,\gamma}$ correlations} 

Liang et al.~(2010) and Lu et al.~(2011) reported the "tight correlations", 
$\Gamma_0\simeq 118\,[E_{iso,\gamma}/10^{52}\, \rm{erg}]^{0.26\pm 0.04}$
and
$\Gamma_0\simeq 264\,[L_{iso,\gamma}/10^{52}\, \rm{erg}]^{0.27\pm 0.03}$
between the initial bulk motion Lorentz factor 
$\Gamma_0$ of the ejecta and $E_{iso,\gamma}$ or the effective luminosity 
$L_{iso,\gamma}= E_{iso,\gamma}/(1+z)\,T_{90} $
where $T_{90}/(1+z)$ is the intrinsic duration 
during which $90\%$ of the observed prompt GRB energy is emitted.
However, $\Gamma_0$ was not a measured value of the initial Lorentz 
factor of the jetted ejecta.
For instance, Liang et al.~(2010) assumed that
\begin{equation}
\Gamma_0\simeq 192\,[E_{iso,52,\gamma}/(t'_p)^3]^{1/8}\,,
\label{fbgam0}
\end{equation}
where $E_{iso,52,\gamma}\!=\!E_{iso,\gamma}/(10^{52}\,{\rm erg})$ 
and $t'_p$ is in seconds. This
relation follows from a fireball model expression for
$\Gamma_0$ \citep{Sari1999} after neglecting its weak dependence on the 
unknown
circumburst density and on the fraction of the relativistic kinetic energy
of the ejecta that is converted to radiation. 

Substitution of
the CB model relations $E_{iso,\gamma}\propto \delta^3$
and  $t'_p \propto 1/(\gamma\,\delta)$ into Eq.~(\ref{fbgam0})
yields $\Gamma_0\propto\gamma^{3/8}\, \delta^{6/8}$.
Consequently,  for a mixed population of  ordinary GRBs and  XRFs,
the CB model predicts the following correlations compared to the observed 
(obs) ones:\\
$\Gamma_0 \propto [E_{iso,\gamma}]^{(5\pm 1)/16}\simeq 
[E_{iso,\gamma}]^{0.31\pm 0.06}$~~~ 
obs: $\Gamma_0\propto [E_{iso,\gamma}]^{0.26\pm 0.04}$,\\ 
$\Gamma_0 \propto [L_{iso,\gamma}]^{(5\pm 1)/16}\simeq 
[E_{iso,\gamma}]^{0.31\pm 0.06}$~~~ 
obs: $\Gamma_0\propto [L_{iso,\gamma}]^{0.27\pm 0.03}$,\\ 
$\Gamma_0 \propto [L_{p,O}]^{(111 \pm 15)/448}
\simeq[L_{p,O}]^{0.25\pm 0.03}$~~~~~ 
obs: $\Gamma_0\propto [L_{p,O}]^{0.20\pm 0.03}$,\\ 
$\Gamma_0 \propto [t'_{p,O}]^{-(21 \pm 3)/32}
\simeq [t'_{p,O}]^{-0.66\pm 0.09}$~~~~~~~ 
obs: $\Gamma_0 \propto [t'_{p,O}]^{-0.59\pm 0.05}$.\\ 
Note that for a GRB population dominated by XRFs and soft GRBs, where the 
dependence on $\gamma$ can be neglected compared to that on $\delta$,
the CB model predicts: 
$\Gamma_0\propto [E_{iso,\gamma}]^{0.25}$, 
$\Gamma_0\propto [L_{iso,\gamma}]^{0.25}$, 
$\Gamma_0 \propto [L_{p,O}]^{0.21}$, and
$\Gamma_0 \propto [t'_p]^{-0.75}$.

\section{Summary and conclusions} 

In a long series of publications we have demonstrated that the cannonball 
model of GRBs predicted correctly the main observed properties of GRBs, 
including the well established correlations between GRB observables, and 
can reproduce successfully the broad band lightcurves of GRBs and their 
afterglows from onset until very late times, despite their enormous 
complexity and diversity \citep[see, e.g.,~][~and references 
therein]{DD2004, DDD2009, DD2009a, DD2009b, DD2010}. In this paper we have 
shown 
that in the CB model, all the newly discovered pair correlations between 
gamma ray burst observables that were reported by Liang et al.~(2010), and 
by Lu et al.~(2011), like the previously well established correlations, 
are a simple consequence of the strong dependence of the GRB observables 
on the Lorentz factor and viewing angle of the highly relativistic and 
narrowly collimated jets whose interaction with the medium along their 
path produces the observed radiations.

One of the newly discovered correlations by Liang et al.~(2010)  
was the tight relationships, $\Gamma_0\!\simeq\! 182 
\,[E_{iso,\gamma}/10^{52}\,{\rm erg}]^{0.25\pm 0.04}$. The authors 
pointed out in their paper that "there is no straight-forward theory that 
predicts this relationship between $\Gamma_0$ and $E_{iso,\gamma}$", and 
that this tight correlation is inconsistent with the well established 
Amati relation \citep{Amati2006} if the prompt gamma ray emission in GRBs 
is produced by the internal shock mechanism of the fireball 
model\footnote{An interpretation of the $\Gamma_0 - E_{\gamma,iso}$ 
correlation discovered by Liang et al.~(2010) was recently proposed by Lu 
et al.~(2011) within the framework of a neutrino cooling dominated central 
engine model.}. However, the tight $\Gamma_0-E_{iso,\gamma}$, correlation 
(and the $\Gamma_0-L_{iso,\gamma}$ correlation discovered by Lu et 
al.~(2011)) were based on values of $\Gamma_0$ inferred from a fireball model 
relation $\Gamma_0 \propto [E_{is0,\gamma}/(t'_p)^3]^{1/8}$ and measured 
values of $E_{iso,\gamma}$ and $t'_p$, and not on reliable measurements of 
the initial Lorentz factor $\gamma$. Using the CB model dependences of 
$E_{iso,\gamma}$ and $t'_p$ on $\gamma$ and $\delta$ one reproduces the 
power-law correlation between $\Gamma_0$ and $E_{iso,\gamma}$ or 
$L_{iso,\gamma}$ with the observed index 0.25 for a population rich in 
soft GRBs and XRFs. Indeed, the GRB sample that was used in Liang et 
al.~(2010) to infer the tight correlation contains a large fraction of 
XRFs (e.g., 060904B, 070318, 070419A, 071010A, 080330, 080710, 070208) and 
soft GRBs. We conclude that the tight correlation between $\Gamma_0$ and 
$E_{iso,\gamma}$ is that expected in the CB model, as well as the Amati 
relation (Eq.~{\ref{Eisomean}), which actually was predicted by the CB 
model \citep[see~][~Eq.~(40)]{DD2000} long before it was discovered 
empirically \citep{Amati2002,Amati2006}. In the case of the standard 
fireball model, where the prompt emission pulses are produced by 
synchrotron emission from internal shocks, the tight 
$\Gamma_0-E_{iso,\gamma}$ correlation yields essentially $E'_{p,\gamma}$ 
that is constant for different $E_{iso,\gamma}$ values, in contradiction 
with the Amati relation \citep{Liang2010}. This, perhaps, is not 
surprising in view of the fact that the standard fireball model was never 
shown to predict correctly the shape of the prompt emission gamma ray 
pulses and their spectral evolution, nor their 
typical photon energy and total GRB energy.

Finally, we would like to caution that the 
new correlations discovered by Liang et al.~(2010)  
were inferred from a  sample of selected GRBs: Only
GRBs with a single optical peak that could be modeled 
with the empirical formula of Kocevski et al.~(2003)
seem to be included in the sample. 
Bright GRBs with clear early time multipeak optical emission were not 
included in the GRB sample. In fact, a single late-time peak can also 
be a late-time flare due to a density bump in the ISM \cite{DDD2002}, 
such as that observed in GRB 970508 \citep{Piro1998},
or the blended sum of unresolved peaks like that observed in XRF 071031 
\citep{Kruhler2009,DD2010}   
Moreover, soft GRBs and XRFs usually have a slow rebrightening 
of their optical and X-ray afterglows that probably has a completely 
different origin - an initial rise in $\delta(t)$ as $\gamma(t)$ 
decreases due to deceleration. Thus, most XRFs and soft GRBs show 
after a  prompt emission flare(s) a slowly rising afterglow 
followed by a  power-law decay like that
of ordinary GRBs \citep{DD2009b}.   

{}

\begin{deluxetable}{lllllc}
\tablewidth{0pt}
\tablecaption{Comparison between the power-law index $k$ of the power-law 
correlations $log[O_i]\!=\!a_{ij}\!+\!k\,log[O_j]$  which was reported
in \citep{Liang2010} for various pairs $(O_i,O_j)$ of GRB observables 
and its value predicted by the CB model.  $t_r,\, t_p,\, t_d$ and $t_w$
stand, respectively, for the rise-time, peak-time, decay-time,  and total 
width of the optical peak.}
\tablehead{
\colhead{Correlated Pair} &\colhead{k [obs]} &\colhead{k [CB model]}
 }
\startdata
$ (t_d,t_r)               $ & $1.06\!\pm\! 0.06$ &               1.00 \\
$ (t_d,t_p)               $ & $1.17\!\pm\! 0.11$ &               1.00 \\
$ (t_r,t_p)               $ & $1.11\!\pm\! 0.08$ &               1.00 \\
$ (t_w,t_p)               $ & $1.16\!\pm\! 0.10$ &               1.00 \\
$ (t_w,t_r)               $ & $1.05\!\pm\! 0.05$ &               1.00 \\
$ (t_w,t_d)               $ & $0.98\!\pm\! 0.01$ &               1.00 \\
$ (L_{p,O},t'_p)          $ & $-2.49\!\pm\! 0.39$&              -2.00 \\
$ (L_{p,O},t'_w)          $ & $-2.00\!\pm\! 0.32$&              -2.00 \\
$ (L_{p,O},E_{iso,\gamma})$ & $1.17\!\pm\! 0.13$ &  $1.25\!\pm\!0.08$ \\      
$( t'_p, E_{iso,\gamma})  $ & $-0.40\!\pm\! 0.07$&  $-0.50\!\pm\!0.17$\\
$(E_{iso,O},E_{iso,\gamma})$ & $0.74\!\pm\! 0.10$&  $ 0.75\!\pm\!0.08$\\ 
$(E'_{p,\gamma}, E_{iso,\gamma})$& $0.51\!\pm\!0.06$& $0.50\!\pm\!0.17$\\    
\hline\\                                               
$(\Gamma_0,E_{iso,\gamma})$ & $0.26 \pm 0.04   $ & $0.31\pm 0.06$\\
$(\Gamma_0,L_{iso,\gamma})$ & $0.27 \pm 0.03   $ & $0.31\pm 0.06$\\
$(\Gamma_0,L_{p,O})$        & $0.20 \pm 0.03   $ & $0.25\pm 0.04$\\
$(\Gamma_0,t'_{p,O})$       & $-0.59\pm 0.05   $ &$-0.66\pm 0.09$\\
\enddata
\label{t1}
\end{deluxetable}

\newpage
\begin{figure}[]
\centering
\vspace{-2cm}
\epsfig{file=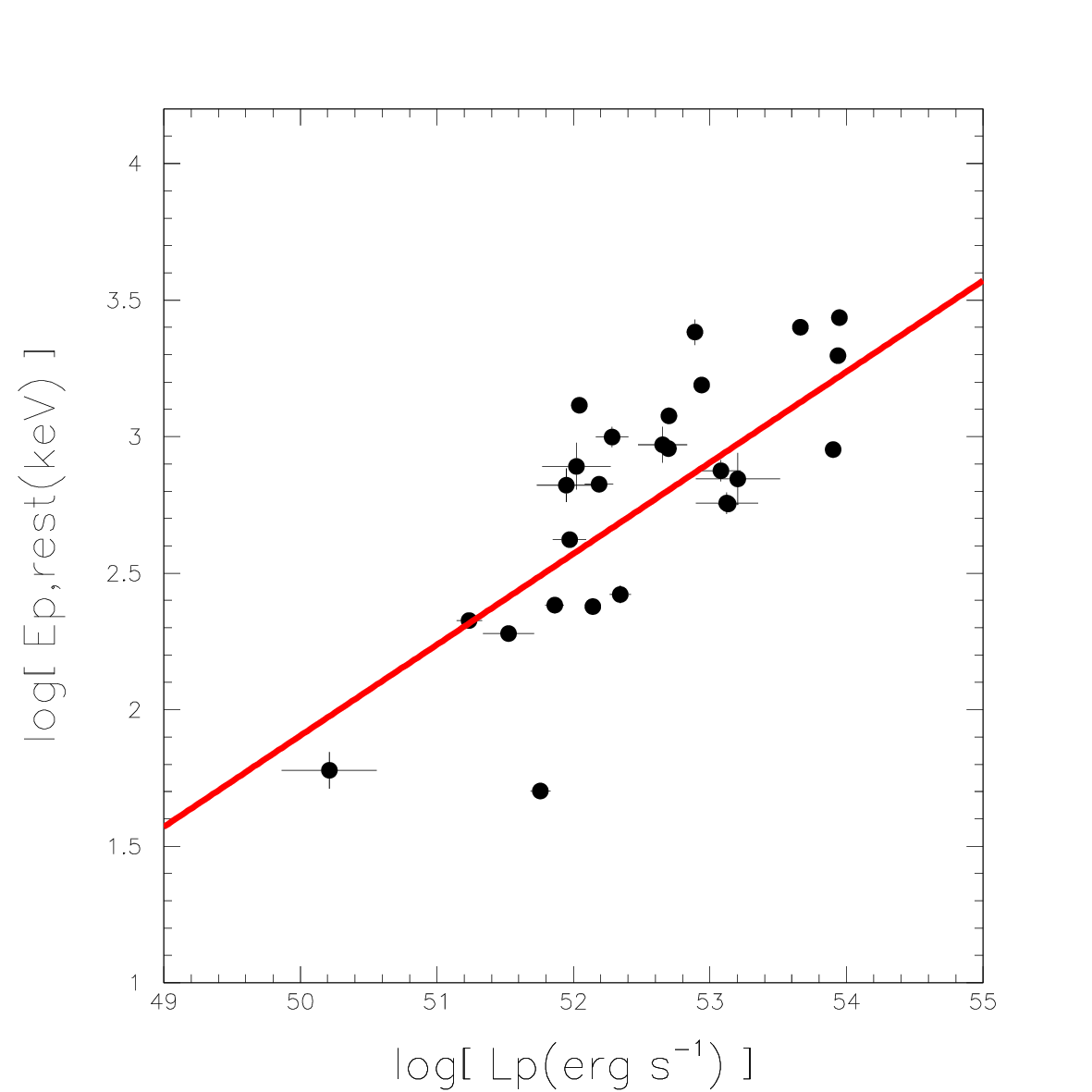,width=12.cm,height=12.cm}
\caption{Comparison between the CB model predicted binary 
correlation $E'_{p,\gamma}\propto [L_{p,\gamma}]^{1/3}$
and the observed correlation inferred for 26 GRBs measured with
Fermi GBM \citep{Gruber2011}.} 
\end{figure}

\end{document}